\documentclass[aps,twocolumn,groupedaddress,showpacs]{revtex4}
\usepackage{tabularx}
\usepackage{graphicx}
\usepackage{dcolumn}
\usepackage{bm}
\def\v_op{ \hat{\mathbf v} }
\def\sigmabar{ \bar{\sigma} }

\newcommand{\ra}{\rangle}

\newcommand{\be}{\begin{equation}}
\newcommand{\ee}{\end{equation}}
\newcommand{\bea}{\begin{eqnarray}}
\newcommand{\eea}{\end{eqnarray}}

\renewcommand{\k}{{\vec{k}}}
\newcommand{\hb}{\hbar}
\renewcommand{\inf}{\infty}

\newcommand{\q}{{\bf {q}}}
\renewcommand{\k}{{\bf{k}}}
\newcommand{\kp}{{\bf{k'}}}

\begin{document}

\title{Auxiliary-field quantum Monte Carlo study of TiO and MnO molecules}

\author{W.~A.~Al-Saidi, Henry Krakauer, and Shiwei Zhang}

\affiliation
{Department of Physics, College of William and Mary,
Williamsburg, VA 23187-8795}

\date{\today}

\begin{abstract}

Calculations of the binding energy of the transition metal oxide
molecules TiO and MnO are presented, using a recently developed
phaseless auxiliary field quantum Monte Carlo approach.  This method
maps the interacting many-body problem onto a linear combination of
non-interacting problems by a complex Hubbard-Stratonovich
transformation, and controls the phase/sign problem with a phaseless
approximation relying on a trial wave function.  It employs random
walks in Slater determinant space to project the ground state of the
system, and allows use of much of the same machinery as in standard
density functional theory calculations, such as planewave basis and
non-local pseudopotentials. The calculations used a single Slater
determinant trial wave function obtained from a density functional
calculation, with no further optimization.  The calculated binding
energies are in good agreement with experiment and with recent
diffusion Monte Carlo results.  Together with previous results for
$sp$-bonded systems, the present study indicates that the phaseless
auxiliary field method is a robust and promising approach
for the study of correlation effects in real
materials.
\end{abstract}

\maketitle

\section{Introduction}

A phaseless auxiliary field (AF) quantum Monte Carlo (QMC) method was
recently introduced \cite{zhang_krakauer} to study correlation effects
in real materials, which has yielded results for a variety of
$sp$-bonded materials in good agreement with experiment and comparable
to those obtained using the standard diffusion Monte Carlo method
(DMC) \cite{QMC_rmp}.  In this paper we present the first application
of the phaseless AF QMC method to the more highly-correlated
transition metal oxide systems.
Because of their complexity (the presence of both localized and 
itinerant characters in the
electronic degrees of freedom, strong electron-ion pseudopotentials,
and the presence of many highly correlated electrons),
there have been relatively few
QMC calculations of any type for transition metal systems
\cite{sokolova,wagner_mitas,lee_mitas_wagner,needs}.

There are many important applications based on the magnetic,
ferroelectric, and superconducting properties of transition metal
oxides. These effects arise from the presence of $d$-shell electrons
whose interactions are often highly correlated.  The generally
successful {\em ab initio\/} density functional theory (DFT) approach
\cite{kohn_nobel} has had limited success in describing these
properties, often predicting incorrect ground states ({\it e.g.\/}
metallic instead of insulating).  Even in cases where correlation
effects are less pronounced and the method is qualitatively correct,
the results are sometimes not of sufficient accuracy.  For example in
ferroelectrics such as PbTiO$_3$, which have essentially no occupied
$d$-states, the relatively small and usually acceptable DFT errors
($\sim$3\%) in predicted equilibrium volumes can lead to suppression
of the ferroelectric ground state.  There is thus a great need for
better theoretical modeling of transition metal systems.

{\em Ab initio\/} quantum Monte Carlo methods are an attractive means
to treat {\em explicitly\/} the interacting many fermion system.
These methods in principle scale algebraically as a low power with
system size.  However, except for a few special cases, QMC methods are
plagued by the fermion sign problem \cite{schmidt84,loh90}, which, if
uncontrolled, results in exponential scaling. No formal solution has
been found for this problem, but approximate methods have been
developed that control it.  The most established QMC method is the
real space fixed-node diffusion Monte Carlo \cite{ceperley_adler}, 
which has been applied to calculate many properties of solids and
molecules \cite{QMC_rmp}. Recent DMC studies have addressed transition
metal systems such as the TiC molecule \cite{sokolova}, TiO and MnO
molecules \cite{wagner_mitas}, solid MnO \cite{lee_mitas_wagner}, and
solid NiO \cite{needs}.

The new phaseless AF QMC approach \cite{zhang_krakauer} is an
alternative that has several appealing features.  For example, it is
formulated in a Hilbert space spanned by some fixed one-particle
basis, and the freedom to choose {\em any\/} one-particle basis
suitable for a given problem could be advantageous.  Moreover, the AF
QMC methodology can take full advantage of well-established techniques
used by independent-particle methods with the same basis. With a
planewave basis, for example, algorithms based on fast Fourier
transforms (FFT) and separable non-local pseudopotentials can be
carried over from DFT planewave codes.  Given the remarkable
development and success of the latter \cite{DFT}, it is clearly
desirable to have a QMC method that can use exactly the same machinery
and systematically include correlation effects by simply building
stochastic ensembles of the independent-particle solutions.

The central idea in standard AF QMC methods \cite{BSS,Koonin} is the
mapping of the interacting many-body problem into a linear combination
of non-interacting problems in external auxiliary fields. Averaging
over different AF configurations is then performed by Monte Carlo (MC)
techniques.  However, except for special cases (e.g., the Hubbard
model with on-site interactions), the two-body interactions will
require auxiliary fields that are {\em complex\/}. As a result, the
single-particle orbitals become complex, and the MC averaging over AF
configurations becomes an integration over complex variables in many
dimensions, and a phase problem occurs.

The phaseless AF QMC method \cite{zhang_krakauer} used in this paper
controls the phase/sign problem in an approximate manner using a trial
wave function. As in fixed-node DMC, the calculated results approach
the exact ones as the trial wave function is improved.  The
ground-state energy in the phaseless method, however, is not a
variational upper bound \cite{zhang_krakauer,Carlson99}.  Previous
results on $sp$-bonded systems \cite{zhang_krakauer,CPC05} and our
current results suggest that the calculated energy is quite
insensitive to the trial wave function.  Accurate ground-state
energies have been obtained with simple trial wave functions, namely
single Slater determinants from DFT or Hartree-Fock calculations.

In this paper, we study the transition metal oxide molecules TiO and
MnO, using the phaseless AF QMC method with planewaves and
pseudopotentials.  This represents the first application of AF-based
QMC to transition metal oxides.  As in regular DFT calculations,
molecules can be studied with planewaves by placing them in large
cells (supercells) and using periodic boundary conditions.  This is
somewhat disadvantageous because one has to insure that the supercells
are large enough to eliminate the spurious interactions between the
images of the molecule. Consequently the computational cost for
isolated atoms and molecules is higher than with a localized basis.
However, the main motivation of the present study is to test the
phaseless AF QMC method for strongly correlated systems such as
transition metal oxides, using the same methodology as previously used
for $sp$-bonded materials.  In addition, a converged planewave basis,
which is straightforward to achieve aside from the computational cost,
gives an unbiased representation of the Hamiltonian, and facilitates
direct comparison with experiment.

The remainder of the paper is organized as follows. The phaseless AF
QMC method is briefly reviewed in Sec.~II. The specific formulation
using a single-particle planewave basis with non-local
pseudopotentials is then discussed in Sec.~III. Finally, in Sec.~IV  we present
results of our calculations for the binding energies of TiO and MnO,
which are in excellent agreement with experiment.

\section{Formalism }

The Hamiltonian for a many-fermion system with two-body interactions
can be written in any one-particle basis in the general form
\begin{equation}
{\hat H} ={\hat H_1} + {\hat H_2}
= \sum_{i,j}^M {T_{ij} c_i^\dagger c_j}
   + {1 \over 2} 
\sum_{i,j,k,l}^M {V_{ijkl} c_i^\dagger c_j^\dagger c_k c_l},
\label{eq:H}
\end{equation}
where $M$ is the size of the chosen one-particle basis, and
$c_i^\dagger$ and $c_i$ are the corresponding creation and
annihilation operators.  Both the one-body ($T_{ij}$) and two-body
matrix elements ($V_{ijkl}$) are known.

As in other QMC methods, the auxiliary field quantum
Monte Carlo obtains  the ground state $\left| \Psi_G \right\rangle$ of
${\hat H}$ by projecting from a trial wave function $\left| \Psi_T
\right\rangle$, using the imaginary-time propagator $e^{-\tau {\hat
H}}$:
\be
\left| \Psi_G \right\rangle \propto  \lim_{n
\to \infty} (e^{-\tau {\hat H}})^n \left| \Psi_T \right\rangle . 
\ee
The trial wave function $\left| \Psi_T \right\rangle$, which should
have  a non-zero overlap with the exact ground state, is assumed to be
in the form of a single Slater determinant or a linear combination of
Slater determinants.

Using a second-order Trotter breakup, we write  
the propagator as:
\begin{equation}
e^{-\tau {\hat H}} = e^{-\tau {\hat H_1}/2} e^{-\tau {\hat H_2}}
e^{-\tau {\hat H_1}/2} + {\cal{O}}(\tau^3).\label{eq:expH}
\end{equation}
The two-body part of the propagator
can be written as an integral of 
one-body operators by a Hubbard-Stratonovich transformation \cite{HS}: 
\begin{equation}
   e^{-\tau{\hat H_2}}
= \prod_\alpha \Bigg({1\over \sqrt{2\pi}}\int_{-\infty}^\infty
            e^{-\frac{1}{2} \sigma_\alpha^2}
           e^{\sqrt{\tau}\,\sigma_\alpha\,
\sqrt{\lambda_\alpha}\,{\hat v_\alpha}} d\sigma_\alpha\Bigg),
\label{eq:HStrans}
\end{equation}
after ${\hat H_2}$ is turned into a sum of squares of one-body operators: 
${\hat H_2} = - {1\over 2}\sum_\alpha
\lambda_\alpha {\hat v_\alpha}^2$, with  $\lambda_\alpha$  a real
number.
 
The propagator of Eq.~(\ref{eq:expH}) can now be expressed 
in the general form:
\begin{equation}
e^{-\tau{\hat H}} =\int P(\sigma) {\cal{B}}(\sigma)\,d\sigma,
\label{eq:prop}
\end{equation}
where we have introduced the  vector representations $\sigma\equiv \{\sigma_1,\sigma_2,
\cdots\}$, $P(\sigma)$ is the normal distribution with mean $0$ and standard deviation $1$, and 
\begin{equation}
{\cal{B}}(\sigma)\equiv
e^{-\tau {\hat H_1}/2}
\,e^{\sqrt{\tau} \sigma\cdot {\hat{\mathbf v}}}
\,e^{-\tau {\hat H_1}/2}, 
\label{eq:Bdef}
\end{equation}
with $\v_op \equiv\{ \sqrt{\lambda_1}\,{\hat v_1},
\sqrt{\lambda_2}\,{\hat v_2}, \cdots\}$.

Monte Carlo methods can be used to evaluate the multi-dimensional
integral of Eq.~(\ref{eq:prop}) efficiently. We follow the 
procedure \cite{Zhang,Zhang_review,zhang_krakauer}
of turning the MC process into an open-ended random walk (instead 
of Metropolis sampling of entire paths along imaginary time \cite{BSS,Koonin}),
because it facilitates the imposition of local constraints
to deal with the sign/phase problem \cite{Zhang_review}.
Each step in the random walk takes 
a Slater determinant $|\phi\ra$ to a new determinant
$|\phi^\prime\ra$:
\be |\phi^\prime (\sigma) \ra =
{\cal{B}}(\sigma) |\phi \ra , \label{eq:proj}
\ee
where $\sigma$ is sampled from $P(\sigma)$. Given sufficient
propagation time one obtains a MC representation of the ground state:
$|\Psi_G\ra \doteq \sum_{\phi} |\phi \ra $.

This straightforward approach, however, will generally lead to
an exponential increase in the statistical fluctuations with the
propagation time.  One can easily understand the source of this by
realizing that the one-body operators $\v_op$ are generally complex,
since $\lambda_\alpha$ usually cannot all be made
positive in Eq.~(\ref{eq:HStrans}).  As a result, the orbitals in
$|\phi\rangle$ will become complex as the propagation proceeds.  This
is the phase problem referred to earlier. It is of the same origin as
the sign problem that occurs when ${\cal B}(\sigma)$ is real. The phase
problem is more severe, however, because for each $|\phi\rangle$,
instead of a $+|\phi\rangle$ and $-|\phi\rangle$ symmetry
\cite{Zhang}, there is now an infinite set $\{ e^{i\theta}
|\phi\rangle\}$ ($\theta \in [0,2\pi)$) from which the MC
sampling cannot distinguish. At large propagation time, the phase of each
$|\phi\rangle$ becomes random, and the MC representation of
$|\Psi_G\rangle$ becomes dominated by noise.

In Ref.~\cite{zhang_krakauer} the phaseless auxiliary field QMC method
was presented to control the phase problem. The first ingredient
of this method is an importance-sampling transformation using a {\em complex}
importance function, $\langle \Psi_T|\phi\rangle$, where
$| \Psi_T \rangle$ is a trial wave function.  In the resulting 
random walk, a walker $|\phi\rangle$
is propagated to a new position 
$|\phi^\prime\rangle$ in each step by 
\begin{equation}
|\phi^\prime(\sigma)\rangle={\cal B}(\sigma-\sigmabar) |\phi\rangle.
\label{eq:prop_imp}
\end{equation}
As in Eq.~(\ref{eq:proj}), $\sigma$ is sampled from $P(\sigma)$, but the
propagator is modified to include a force bias, or 
shift \cite{shiftcont_rom97}: 
\begin{equation}
\sigmabar =
- \sqrt{\tau}
{\langle\Psi_T|\v_op|\phi\rangle \over \langle\Psi_T | \phi\rangle}.
\label{eq:FB}
\end{equation}
A walker  carries a weight
$w_{\phi}$ which is updated according to
\begin{equation}
w_{\phi^\prime}=W(\phi)\,w_\phi, 
\label{eq:wt_imp}
\end{equation}
where $W(\phi)$ can 
be expressed in terms of the so-called local energy, $E_L$:
\begin{equation}
W(\phi) \doteq 
\exp\bigg[-\tau  
{\langle\Psi_T|\hat{H}|\phi\rangle \over \langle\Psi_T | \phi\rangle}\bigg]
\equiv \exp[-\tau E_L(\phi)].
\label{eq:El}
\end{equation}

In the limit of an exact $|\Psi_T\rangle$, $E_L$ is a real
constant, the weight of each walker remains real, and the 
mixed estimate for the energy is phaseless:
\begin{equation}
E_G = 
{\langle\Psi_T|\hat{H}|\Psi_G\rangle \over \langle\Psi_T | \Psi_G\rangle}
\doteq
{\sum_{\phi^\prime} w_{\phi^\prime} E_L({\phi^\prime}) 
\over 
\sum_{\phi^\prime} w_{\phi^\prime}}.
\label{eq:mixed_w_EL}
\end{equation}
With a general $|\Psi_T\rangle$ which is not exact, a natural
approximation is to replace $E_L$ in Eq.'s (\ref{eq:El}) and
(\ref{eq:mixed_w_EL}) by its real part, ${\rm Re}\, E_L$,  
leading to a phaseless formalism for the random walk, with real and
positive weights.

The second ingredient in the phaseless method involves a projection:
the modified random walk is still ``rotationally invariant'' in the
complex plane defined by $\langle\Psi_T|\phi\rangle$. With the
propagation, the walkers will populate the complex plane symmetrically
independent of their initial positions. In particular, a finite
density of walkers will develop at the origin where the local energy
$E_L(\phi)$ diverges, and this causes diverging fluctuations in the
weights of walkers.

This problem, which is inherent of the ``two-dimensional'' nature of
the random walk in the complex plane, can be controlled with an
additional approximation, in which the random walk is projected to
``one-dimension.'' This is done, e.g., by multiplying the weight of
each walker in each step by $\max\{0,\cos(\Delta\theta)\}$, where
$\Delta\theta$ is the phase of
$\langle\Psi_T|\phi^\prime\rangle/\langle\Psi_T |\phi\rangle$.  The
projection ensures that the density of walkers vanish at the origin.
Note that the projection has no effect 
when $\v_op$ is real.  This additional approximation and the
importance-sampling procedures of Eq.'s (\ref{eq:prop_imp}) through
(\ref{eq:El}) form the basis of the new phaseless AF QMC method.

\section{Implementation with Planewaves}

The calculations reported in this paper were carried out in supercells using a
planewave basis and periodic boundary conditions (PBC).
Pseudopotentials are used as in DFT calculations to
represent the electron-ion interaction, eliminating the core electrons 
from the Hamiltonian.  The basis set consists of the $M$
planewaves with kinetic energy $|\k|^2/2 < E_{{\rm{cut}}}$, where the
parameter $E_{{\rm{cut}}}$ is a cutoff energy.

In a planewave basis, the one-body operator $\hat H_1$ of Eq.~(\ref{eq:H})
is the sum of the kinetic energy and the electron-ion
interaction, and $\hat H_2$ represents the electron-electron
interaction. These can be expressed as:
\bea 
\hat H_1&=& -\frac{\hb^2}{2m}\sum_{\k,s} |\k|^2
c^{\dag}_{\k,s} c_{\k,s} 
+ \sum_{\k,\kp,s} V_{L}(\k-\kp) c^{\dag}_{\k,s} c_{\kp,s} \nonumber \\
&\,\,\,\,\, +& \sum_{\k,\kp,s} V_{NL}(\k,\kp) c^{\dag}_{\k,s}
c_{\kp,s} \nonumber \\
\hat H_2 &=& \frac{1}{2 \Omega} \sum_{\k,\kp,s,s'}
\sum_{\q \neq 0}
 \frac{4 \,\pi\, e^2}{|\q|^2} \, c^{\dag}_{\k+\q,s}
c^{\dag}_{\kp-\q,s'} c_{\kp,s'} c_{\k,s}. \nonumber\\ 
\eea
Here $c^{\dag}_{\k,s}$ and $c_{\k,s}$ are the creation and
annihilation operators of an electron with momentum $\k$ and spin
$s$. 
$V_L(\k-\k')$ and $V_{NL}(\k,\k')$ are the local and non-local parts of the 
pseudopotential, respectively.
$\Omega$ is the super-cell volume, $\k$ and $\kp$ are planewaves
within the cutoff radius, and the $\q$-vectors
satisfy $|\k+\q|^2/2 < E_{\rm{cut}}$.

A Hubbard-Stratonovich transformation is applied to decouple the
electron-electron interaction $\hat H_2$ into a linear combination of
one-body operators.  The resulting one-body operators consist of
density operators of the form $\hat \rho(\q)=\sum_{\k,s}
c^{\dag}_{\k+\q,s} c_{\k,s}$.  The number of auxiliary fields is
proportional to the number of unique $\q$ vectors that the basis
allows, i.e., roughly eight times the number of planewaves in the
basis.

\begin{table}
\caption{A summary of the binding energy BE (in eV), equilibrium bond
  length $R_{e}$ (in a.u.) and harmonic vibrational frequency $\omega$
  (in cm$^{-1}$) of the TiO molecule with two different
  pseudopotentials. The first, with $E_{\rm cut}=50\,$Ry ($50\,$Ry
  psp), was used in all ensuing DFT and QMC calculations. The second
  has a $64\,$Ry cut-off.  The corresponding values of the cut-off
  radius, $r_c$, are listed in the footnotes (in units of a.u.).  DFT
  results from both Perdew-Burke-Ernzerhof (PBE) GGA \cite{gga} and
  Perdew-Wang 92 LDA \cite{lda} functionals are shown, together with
  experimental values.}
\begin{tabularx}{2.88 in }{r  c c c }
\hline
\hline
         &   BE  &  $R_{e}$ & $\omega$      \\ 
\hline
Experiment\cite{TiO_ref_exp1,TiO_ref_exp2}  \qquad \qquad &  6.87 or 6.98  & 3.06  & 1009  \\
\hline
$50\,$Ry psp\footnote{O $r_{c}$: $1.45$ ($s$), $1.55$ ($p$); 
Ti $r_{c}$: $1.40$ ($s$),  $1.40$ ($p$),  $1.80$ ($d$).}
\ \ \ GGA \qquad \qquad & 8.00  & 3.02  & 1005  \\ 
 LDA \qquad \qquad & 9.11  & 2.99  & 1040  \\
\hline
$64\,$Ry psp\footnote{O $r_{c}$: $1.30$ ($s$), $1.39$ ($p$);
Ti $r_{c}$: $1.35$ ($s$),  $1.35$ ($p$), $1.52$ ($d$).}
\ \ \ GGA \qquad \qquad & 7.96  &  3.04  & 1008 \\
LDA \qquad \qquad & 9.05  &  3.02  &     1044   \\
\hline
\hline
\end{tabularx}
\label{table_dft_TiO}
\end{table}

Non-local pseudopotentials can be treated {\em exactly} within the
present AF QMC formalism, and the use of separable forms leads to the
same speed-up achieved in planewave DFT calculations
\cite{zhang_krakauer}. This is to be compared with standard real-space
DMC calculations where an additional locality approximation
\cite{Mitas91} is used for non-local pseudopotentials  that depends on
the overall quality of the trial wave function $|\Psi_T\rangle$. (In
contrast, the fixed-node approximation in DMC only depends on the
position of the nodal surface of $|\Psi_T\rangle$.) In order to
minimize errors due to the locality approximation, small
pseudopotential cut-off radii $r_c$ tend to be used.  This could
result in harder pseudopotentials than otherwise required by
transferability considerations. In the AF QMC, the use of non-local
pseudopotentials with larger values $r_c$ (determined only by
transferability requirements) does not pose any additional difficulty.

\section{Results}

\begin{table}
\caption{A summary of the binding energy BE (in eV), equilibrium
  bond length $R_{e}$ (in a.u.), and harmonic vibrational frequency $\omega$  
(in cm$^{-1}$) of the MnO molecule with three different
  pseudopotentials. The first, with $E_{\rm cut}=64\,$Ry and
created from the design non-local (DNL) procedure, was used in all
  ensuing DFT and QMC calculations.  Two
other sets are also tested here, with $64\,$Ry and $82\,$Ry cut-off values
and without DNL. The corresponding $r_c$ values (in a.u.)
are listed in the footnotes. 
Calculated results are from DFT GGA.}
\begin{tabularx}{2.4 in }{ p{1.5 in} c  c c }
\hline
\hline
      &   BE  &  $R_{e}$ & $\omega$      \\ 
\hline
Experiment~\cite{TiO_ref_exp2}   &  3.70   & 3.11  & 832  \\
\hline
$64\,$Ry DNL psp\footnote{O $r_{c}$: $1.45$ ($s$), $1.55$ ($p$);
Mn $r_c$:  $1.40$ ($s$),
   $1.40$ ($p$), $1.65$ ($d$).
}  &  5.11  &  3.11  & 822  \\
$64\,$Ry psp\footnote{O $r_{c}$: $1.45$ ($s$), $1.55$ ($p$);
Mn $r_c$:  $1.40$ ($s$),
   $1.40$ ($p$), $1.65$ ($d$).}& 4.90  &  3.07  & 878  \\
$82\,$Ry psp\footnote{O $r_{c}$:
$1.05$  ($s$), $1.02$ ($p$);
Mn $r_c$:  $1.25$ ($s$),
   $1.25$ ($p$), $1.50$ ($d$).}  & 4.99 &  3.09 & 845  \\
\hline\hline
\end{tabularx}
\label{table_dft_MnO}
\end{table}

In this paper, we apply the phaseless AF QMC method to calculate the
binding energies of the transition metal oxide molecules TiO and MnO.
Norm-conserving pseudopotentials are used, and the non-local part of
the pseudopotential $V_{NL}$ is represented using the separable
Kleinman-Bylander (KB) form \cite{KB-separable}.

To obtain the trial wave function $|\Psi_T\rangle$ for each QMC
calculation, a DFT calculation with the generalized gradient
approximation (GGA)  is carried out with the ABINIT \cite{Abinit}
program, using the same pseudopotentials and planewave basis.
$|\Psi_T\rangle$ is then taken as the single Slater determinant formed
from the occupied single-particle orbitals obtained from this DFT
calculation, with {\em no further optimization}.  The random walkers
are all initialized to $|\Psi_T\rangle$, so the many-body ground-state
projection initiates from the GGA state.  In addition, this
$|\Psi_T\rangle$ is used in the QMC calculations to control the
sign/phase problem as described in Section~II.

The pseudopotentials were generated by the OPIUM program \cite{rappe}
using Ti$^{++}$, Mn$^{++}$, and neutral oxygen as reference
configurations.  The titanium and manganese semicore states
(3s$^2$3p$^6$) were included as valence states, so the Ti and Mn atoms
contribute 12 and 15 valence electrons, respectively, while the O atom
contributes 6 electrons.

Well-converged planewave cutoffs were $50\,$Ry for oxygen and
titanium, and $64\,$Ry for manganese.  These $E_{\rm cut}$'s were
chosen such that the resulting cut-off errors, systematically analyzed
using DFT calculations, were much smaller than the expected QMC
statistical errors. In addition, we have carried out QMC calculations
on a $1\times 1 \times 1$ TiO solid supercell with a 50~Ry and 60~Ry
cutoff, respectively. The calculated energies are the same within
statistical error bars ($\approx 0.1$~eV), confirming basis
convergence at the correlated level.  The Mn pseudopotential is
created using the design non-local pseudopotential procedure
\cite{rappe2}. This enhances the pseudopotential transferability by
exploiting the flexibility contained in the separable KB form of the
nonlocal pseudopotential.

The accuracy of the pseudopotentials was examined with DFT
calculations of binding energies, as well as the equilibrium bond
length and harmonic vibrational frequencies. In Tables
\ref{table_dft_TiO} and \ref{table_dft_MnO}, we summarize our
calculations of these properties for different OPIUM pseudopotentials.
In both cases, increasing the hardness of our pseudopotentials did not
lead to significant changes in the calculated properties.  We have
also done some of these calculations using
Troullier-Martins~\cite{trouillier_martin} pseudopotentials with the
same cutoff radii, and little difference was found. Moreover, our LDA
results for the bondlengths for TiO and MnO, $R_{\rm{e}}=
2.99$~a.u. and $3.05$~a.u., are  in reasonable agreement with the
all-electron  LDA values ($R_{\rm{e}}= 3.020$~a.u. and
$3.032$~a.u.) \cite{Hartwigsen} and those obtained with the
Hartwigsen-Goedecker-Hutter pseudopotentials \cite{Hartwigsen}. 
The TiO results of
$R_e$ and $\omega$ also compare favorably with the calculations of
Ref. \cite{albaret}.

\begin{table}
\caption{ A comparison between LAPW and pseudopotential calculations
  for  non spin-polarized TiO in a $7 \times 7 \times 14$~a.u.$^3$
  supercell.  We show the 
equilibrium  bond length $R_{e}$ (in a.u.) and harmonic vibrational frequency $\omega$  (in
  cm$^{-1}$) from DFT, using both GGA and LDA. The two OPIUM pseudopotentials
are the same as those in Table \ref{table_dft_TiO}.
}
\begin{tabularx}{2.4 in }{l p {0.9 in} c c  }
\hline
\hline
      &          &   $R_{e}$ & $\omega$      \\ 
\hline
LAPW & GGA  \qquad\qquad    &  3.01   & 1057 \\
     & LDA      &  2.97    &  1097 \\
\hline
$50\,$Ry psp \quad \quad & GGA
&  2.96   & 1060  \\
     & LDA   &  2.94   & 1095  \\
\hline
$64\,$Ry psp & GGA 
& 2.99 & 1058 \\  
     & LDA   & 2.97 & 1091 \\  
\hline\hline
\end{tabularx}
\label{table_LAPW_TiO}
\end{table}

As a further check on the pseudopotentials, we have carried out a
comparison between pseudopotential and all-electron
LAPW calculations. The latter is computationally more costly, so we
limited the comparison to a $7 \times 7 \times 14$~a.u.$^3$
supercell for non spin-polarized TiO molecule.  Our results for the
calculated equilibrium bond length and angular frequency of vibration
are summarized in Table~\ref{table_LAPW_TiO}. The close agreement
between the LAPW and the pseudopotential results gives further
evidence on the reliability of the pseudopotentials.

Clearly these tests on the quality of the pseudopotentials are far
from perfect. Our pseudopotentials are all DFT-based, and the tests
are with DFT calculations. For $sp$-bonded systems, we have done
plane-wave Hartree-Fock (HF) calculations using OPIUM DFT pseudopotentials,
and compared with all-electron HF results. In general, these tend to
be quite consistent with the DFT tests, and often good agreement at
the HF level is found when good test results have been obtained from
DFT calculations. Of course, the 
suitability of a DFT or HF pseudopotential (i.e., derived from
independent-particle procedures) for many-body calculations is a 
separate issue, which our tests do not address. 
Empirically, such pseudopotentials have been widely used in many-body 
calculations and have been quite successful.

\begin{table}
\caption{ A summary of the calculated binding energy of the molecule
  TiO for different supercells.  Supercell dimensions are given in
  a.u.~and binding energies are in eV. The QMC statistical errors are
  in the last two digits, and are indicated in parentheses.  At the
  DFT GGA level, the binding energy converges to 8.00\,eV.
}
\begin{tabularx}{3 in}{p {1.9 in} c c}
\hline
\hline
                             &  GGA       &  QMC      \\
\hline
10$\times$11$\times$17       &  7.46     &  6.59(20)  \\
12$\times$12$\times$15       &  7.77     &  6.98(21)   \\
14$\times$14$\times$15       &  7.94     &  7.08(21)   \\
$\inf$                       &  8.00     &             \\
\hline\hline
\end{tabularx}
\label{table_BE_TiO}
\end{table}

\begin{table}
\caption{The calculated total ground-state energy of Mn for different
  supercells. Supercell dimensions are in a.u.~and energies are in eV.
  The QMC statistical errors are in the last  digit, and are shown in
  parentheses.}
\begin{tabularx}{3.1 in}{p {1.6 in} c c }
\hline
\hline
                                           &  GGA           &  QMC     \\
\hline
11$\times$12$\times$15          \,\,\,\,   & -2766.66  \,\,\,\,        & -2766.40(5)    \\
12.55$\times$13.69$\times$17.11 \,\,\,\,   & -2766.38  \,\,\,\,        & -2765.66(4)    \\
14$\times$14$\times$15          \,\,\,\,   & -2766.32  \,\,\,\,        & -2765.89(9)    \\
15.4$\times$15.4$\times$16.5    \,\,\,\,   & -2766.25  \,\,\,\,        & -2765.74(8)    \\
\protect{$\inf$}                \,\,\,\,   & -2766.20  \,\,\,\,        &                \\
\hline\hline
\end{tabularx}
\label{table_Mn_cells}
\end{table}

The use of PBC with a planewave basis requires supercells that
are large enough to control spurious interactions between the periodic
images of the system under study. We studied convergence with respect to such
size effects using both ABINIT and QMC calculations.
Representative results are
shown in Tables~\ref{table_BE_TiO} and \ref{table_Mn_cells}.

Estimating the size-effects in the AF QMC calculations is complicated
by the presence of finite Totter time-step ($\tau$) errors. The QMC
values shown in Tables~\ref{table_BE_TiO} and \ref{table_Mn_cells} are
final values after extrapolations in $\tau$, as discussed further
below.  The range of supercells shown in Table~\ref{table_BE_TiO}
corresponds to about 12,000-17,000 planewaves in our basis. For the Ti
atom, the largest two supercells resulted in a degeneracy of the
highest-lying occupied $d$-orbitals in the density functional
calculations.  To break the degeneracy, these supercells were modified
to $11.6 \times 12\times 15$~a.u.$^3$ and $13.5 \times 14 \times 15$~
a.u.$^3$, respectively. The fully converged value of the DFT GGA TiO
binding energy is 8.00\,eV, as shown.  For the AF QMC calculations,
the binding energies for the larger sizes are converged to well within
the statistical errors.

Table~\ref{table_Mn_cells} shows the energy of the Mn atom
for different supercell sizes. The corresponding number of planewaves
is between 17,000 and 34,000.
As can be seen, the QMC energy is converged to less than the
statistical error for the $14\times 14\times 15$ supercell, although 
for the smaller supercells, the finite-size errors are significant both 
in GGA and QMC. The MnO molecule, on the
other hand, exhibits a much smaller size effect, with QMC
energies of $-3195.50(11)$\,eV and $-3195.58(7)$\,eV 
for the $11\times12\times15$ and $14\times14\times15$ supercells,
respectively.

\begin{figure}
\includegraphics[width=0.45\textwidth]{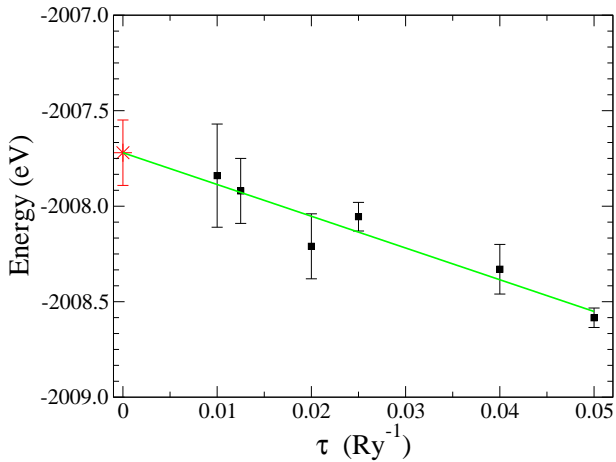} 
\caption{QMC time-step $\tau$ dependence of the total energy of the
TiO molecule. A $10\times 11\times17$~a.u.$^3$ supercell was used. The
solid line is a linear fit to the calculated QMC energies (solid
squares). The final extrapolated energy $E=-2007.72(17)$\,eV is shown
as a star. }
\label{fig_TiO}
\end{figure}

\begin{figure}[htb]
\centering
 \includegraphics[width=0.5\textwidth]{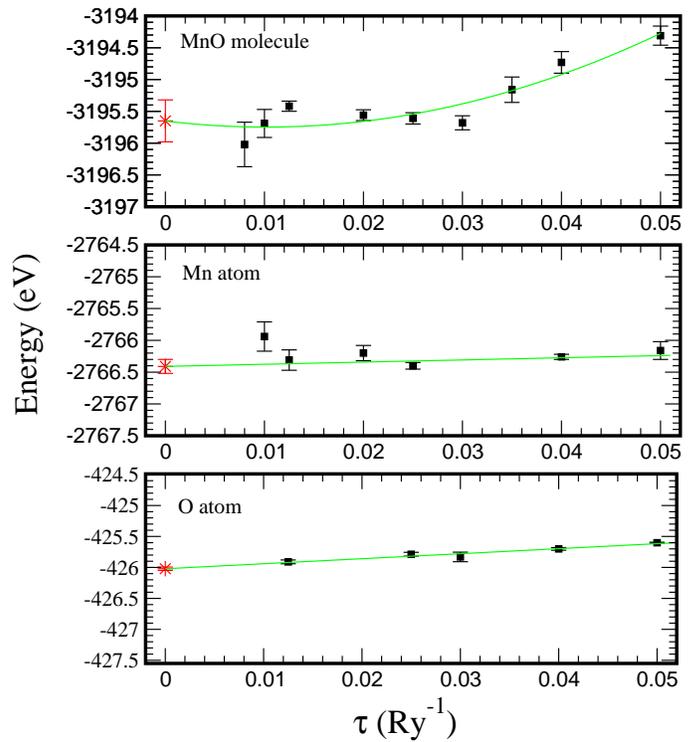}
\caption{QMC time-step $\tau$ dependence of MnO, Mn, and O.
An $11\times 12\times15$\,a.u.$^3$ supercell was used for MnO and Mn, 
and a $10\times
11\times17$\,a.u.$^3$ supercell for oxygen. The solid line is a least-squares
fit to
the QMC energies (solid squares). The final extrapolated values 
are shown as a star. MC statistical error bars are indicated.
}
\label{fig_MnO}
\end{figure}

The QMC Trotter errors were examined by studying the individual
time-step dependence for the atoms and the molecule using a particular
supercell size. Figure~\ref{fig_TiO}, for example, shows the Trotter
extrapolation for the TiO molecule, done with a $10\times11\times 17$
supercell.  The Trotter behavior obtained from this procedure was then
used to extrapolate the QMC data of other supercell sizes, for which
calculations were performed with the time step fixed at
$\tau=0.025~\rm{Ry}^{-1}$.  The final extrapolated results are what is
shown (e.g., in Table~\ref{table_BE_TiO}).  Figure~\ref{fig_MnO}
shows the time-step dependence of MnO, which exhibits a
quadratic behavior compared to the more linear dependence in
Fig.~\ref{fig_TiO} for TiO.  The Mn and O atoms exhibit much smaller
finite-$\tau$ errors, as is also the case with the  Ti atom (not shown).

\begin{table}
\caption{A summary of the binding energies of the molecules TiO, MnO,
  and O$_2$. Calculated results from the present QMC method and
  diffusion Monte Carlo (TiO and MnO from Ref.~\cite{wagner_mitas},
  and O$_2$ from Ref.~\cite{grossman}) are shown, together with
  experimental values (TiO from
  Refs.~\cite{TiO_ref_exp1,TiO_ref_exp2}, MnO from
  Ref.~\cite{TiO_ref_exp1}, and O$_2$ from Ref.~\cite{grossman}).
  Equilibrium experimental bond lengths were used in the molecule
  calculations.  Our QMC used as trial wave function a single Slater
  determinant from DFT GGA. The trial wave functions used in DMC are
  indicated in the footnotes.  All energies are in eV, and the
  experimental zero point energy is added to each molecule.  }
\begin{tabularx}{3 in }{ p {1.3 in} l l l }
\hline
\hline
      & TiO & MnO  & O$_2$  \\ 
\hline
Experiment  & 6.98   &
3.70  & 5.1152(9)  \\
   & 6.87  & &   \\
\hline
DMC (HF)\footnote{Trial w.f. is a  (HF 1-determinant)$\times$Jastrow.} & 6.3(1)  & 2.9(1) & 4.84(2)\\
DMC (B3LYP)\footnote{Trial w.f. is a (DFT B3LYP 1-determinant)$\times$Jastrow.}  & 6.9(1) & 3.4(2) &  \\
DMC (MCSCF)\footnote{Trial w.f. is a 
  (MCSCF multi-determinant)$\times$Jastrow.}  & 6.7(2)  & 3.4(2) &
\\
\hline
Present QMC    & 7.02(21) &   3.79(34)   & 5.12(10)\\   
\hline\hline
\end{tabularx}
\label{table_final}
\end{table}

Table~\ref{table_final} summarizes the results for the molecular
binding energies. For comparison we also include results from a recent
diffusion Monte Carlo study by Wagner and Mitas \cite{wagner_mitas}.
As mentioned, our AF QMC calculations use a single-determinant trial
wave functions obtained from a DFT GGA calculation, without a Jastrow
factor or any further optimization to the determinant.  We see that
the calculated binding energies from AF QMC and those from DMC
\cite{wagner_mitas} with trial wave functions containing either an 
optimized hybrid B3LYP determinant or multiple determinants from MCSCF
are in good agreement with each other and with experiment. DMC with a
trial wave function containing the Jastrow and a single Slater
determinant from HF, on the other hand, gives somewhat worse agreement
with experiment. We have not carried out AF QMC calculations using an
HF trial wave function for these molecules. In several $sp$-bonded
molecules, DFT and HF-generated trial wave functions showed little
difference in the calculated energies in AF QMC.

We have also included in Table~\ref{table_final} the results for the
binding energy of the O$_2$ molecule.  Because of the short bond
length of this molecule (R$_e$=2.281~a.u.), a harder pseudopotential
was used, with a higher $E_{\rm cut}$ of $82$\,Ry and smaller values
for $r_c$ (last entry in Table \ref{table_dft_MnO}).  At the DFT GGA
level the binding energy is $5.72$~eV. Our QMC results shown in
Table~\ref{table_final} were obtained using a supercell of size $8
\times 9 \times 11$~a.u.$^3$. Additional QMC and DFT calculations with
a larger supercell of $11 \times 12 \times 13$~a.u.$^3$ have verified
that the finite-size effects are within our statistical error bars
($\approx 0.1$~eV). Again, we see that the agreement with experiment
is very good.

Finally, we comment briefly on the computational cost.  As mentioned,
the use of planewaves for isolated molecules is somewhat
disadvantageous even at the density-functional level, because of the
need for large supercell sizes to reduce the spurious interactions
between the images of the molecule. The number of planewaves, $M$, is
proportional to the supercell volume, and the computational cost
scales with $M$ as $ M \ln M $. (In addition, it scales quadratically
with the number of electrons.)  As a result, these planewave AF QMC
calculations are computationally rather demanding, especially with
transition metal oxides. For instance, the ground-state energy of the
MnO molecule in Fig.~2 at the single Trotter step of $\tau =
0.008\,\rm{Ry}^{-1}$ (with an error bar of $0.35$\,eV) was obtained
from running on an Intel XEON cluster (3.2 GHz) for about 150 hours
using 72 processors.

In summary, we have presented the first study of transition metal
oxide molecules by AF QMC. We have shown that the binding energies of
TiO and MnO calculated with the new phaseless AF QMC method
\cite{zhang_krakauer} are in good agreement with experiment, and are
comparable to the best results obtained from diffusion Monte Carlo
\cite{wagner_mitas}. It is encouraging that a trial wave function of
only DFT single Slater determinants was sufficient for the phaseless
QMC method to reach this accuracy. Together with previous results for
$sp$-bonded systems \cite{zhang_krakauer,CPC05}, the present study
indicates that the phaseless method is a robust QMC
method. Complementary to standard DMC, it offers a promising approach
for the computation of correlation effects in real materials.

\section{Acknowledgments:}

We would like to thank E.~J.~Walter for many useful discussions, and
for carrying out the LAPW calculations.  This work was supported by
ONR Grants N000149710049 and N000140110365, NSF, and DOE's
Computational Materials Science Network. 
Computations were carried out in part at the Center for Piezoelectrics by
Design (CPD), the SciClone Cluster at the College of William and Mary,
the National Center for Supercomputing Applications (NCSA), and the San
Diego Supercomputer Center (SDSC) center.

\end{document}